# Electronic- and band-structure evolution in low-doped (Ga,Mn)As

O. Yastrubchak [1]*, J. Sadowski [2,3], H. Krzyżanowska [1,4], L. Gluba [1], J. Żuk [1], J.Z. Domagala [3], T. Andrearczyk [3], and T. Wosinski [3]

[1] *Institute of Physics, UMCS, Pl. Marii Curie-Skłodowskiej 1, 20-031 Lublin, Poland*
[2] *MAX-IV Laboratory, Lund University, P.O. Box 118, SE-221 00 Lund, Sweden*
[3] *Institute of Physics, Polish Academy of Sciences, 02-668 Warszawa, Poland*
[4] *Department of Physics and Astronomy, Vanderbilt University, 6506 Stevenson Center, Nashville, TN 37325, USA*

*E-mail: yastrub@hektor.umcs.lublin.pl*

Abstract

Modulation photoreflectance spectroscopy and Raman spectroscopy have been applied to study the electronic- and band-structure evolution in (Ga,Mn)As epitaxial layers with increasing Mn doping in the range of low Mn content, up to 1.2%. Structural and magnetic properties of the layers were characterized with high-resolution X-ray diffractometry and SQUID magnetometery, respectively. The revealed results of decrease in the band-gap-transition energy with increasing Mn content in very low-doped (Ga,Mn)As layers with *n*-type conductivity are interpreted as a result of merging the Mn-related impurity band with the host GaAs valence band. On the other hand, an increase in the band-gap-transition energy with increasing Mn content in (Ga,Mn)As layers with higher Mn content and *p*-type conductivity indicates the Moss-Burstein shift of the absorption edge due to the Fermi level location within the valence band, determined by the free-hole concentration. The experimental results are consistent with the valence-band origin of mobile holes mediated ferromagnetic ordering in the (Ga,Mn)As diluted ferromagnetic semiconductor.

# I. INTRODUCTION

Ferromagnetic semiconductors, integrating both semiconducting and magnetic properties, are promising materials for new class spintronic devices. Among those materials, (Ga,Mn)As has become a model diluted ferromagnetic semiconductor widely used for testing novel device functionalities.[1] Homogeneous (Ga,Mn)As layers containing up to about 10% of Mn atoms can be grown by low-temperature (200–250°C) molecular-beam epitaxy (LT-MBE).[2] Mn atoms substituting the Ga lattice sites in the GaAs matrix, $Mn_{Ga}$, act as acceptors supplying both mobile holes and magnetic moments. Below the Curie temperature, $T_C$, substitutional $Mn^{2+}$ ions, with half-filled $3d$ shell and $S = 5/2$ spin moment, are ferromagnetically ordered owing to interactions with spin-polarized holes. $Mn_{Ga}$ acceptors, with an impurity binding energy of 0.11 eV,[3,4] are more localized than shallower, hydrogenic-like acceptors in GaAs. This results in a higher critical carrier density for the metal-insulator transition (MIT) in Mn doped GaAs, of about $1\times10^{20}$ cm$^{-3}$,[5,6] as compared to the critical density for the shallow acceptors in GaAs, which is two orders of magnitude lower. Importantly, homogeneous ferromagnetic ordering in (Ga,Mn)As occurs on both the insulating and metallic side of the MIT.[7] However, the character of electronic states near the Fermi energy and the valence-band structure in this material are still a matter of controversy.[6,8] There are two alternative models of the band structure in (Ga,Mn)As. The first one involves persistence of the $Mn_{Ga}$-related impurity band on the metallic side of the MIT with the Fermi level residing within the impurity band and mobile holes retaining the impurity band character.[9-12] The second one assumes mobile holes residing in nearly unperturbed valence band of the GaAs host, which play a key role in the *p-d* Zener model of ferromagnetism in diluted magnetic semiconductors.[13,14] In this model the Fermi level position is determined by the concentration of valence-band holes.

In order to resolve the above controversy, we have recently investigated fundamental properties of the energy band structure in ferromagnetic (Ga,Mn)As epitaxial layers, with Mn content in the range from 1% to 6%, by using modulation photoreflectance (PR) spectroscopy.[15] From analysis of the PR spectra we obtained the interband transition energies in (Ga,Mn)As layers with various Mn contents. In (Ga,Mn)As with a low (up to 2%) Mn content this energy was slightly increased with respect to that in the reference GaAs layer, which was interpreted as a result of the Moss-Burstein shift of the absorption edge due to the Fermi level location below the top of the GaAs valence band,[16] in agreement with the Zener model. On the other hand, in (Ga,Mn)As with higher Mn content we revealed significant decrease in the band-gap-transition energy with increasing Mn content. We interpreted those results in terms of a disordered valence band, extended within the band gap, formed, in highly Mn-doped (Ga,Mn)As, as a result of merging the $Mn_{Ga}$-related impurity band with the host GaAs valence band, in agreement with the band structure proposed by Jungwirth et al.[17,18] Our experimental results are also consistent with recent first-principles calculations of the (Ga,Mn)As band-gap energy[19] and the very recent experimental results of hard X-ray angle-resolved photoemission spectroscopy.[20] The disordered character of the valence band may account for the observed very low mobility of itinerant holes in ferromagnetic (Ga,Mn)As layers.

In this paper we report results of modulation PR spectroscopy investigations of (Ga,Mn)As layers with a very low Mn content, in the range 0–1.2%, i.e. in the range where the onset of ferromagnetic ordering occurs at the Mn content of about 1%. Initially undoped LT-MBE-grown GaAs (LT-GaAs) layers are *n*-type with a rather low free-electron concentration.[21] During the growth a significant amount of about 1% excess arsenic is incorporated into the GaAs matrix, mainly in a form of arsenic antisites, $As_{Ga}$, with a typical concentration of $1\times10^{20}$ cm$^{-3}$, which determine the electronic properties of LT-GaAs.[21,22] The $As_{Ga}$ defects act as double donors in GaAs giving rise to two deep energy levels in the band gap: the single donor level at $E_C - 0.75$ eV and the double donor level at $E_V + 0.52$ eV,[23] where $E_C$ and $E_V$ are the conduction and valence band edges, respectively. When Mn is substitutionally introduced in the LT-GaAs host, the $As_{Ga}$ levels become depopulated resulting in compensation of the $Mn_{Ga}$ acceptors in very dilute (Ga,Mn)As layers. Only at higher Mn content, of about 1%, the layers become *p*-type with a sufficient concentration of mobile holes in order to create the ferromagnetic ordering at low temperatures through their interaction with the Mn local moments.

## II. EXPERIMENTAL

In the present work we investigated a series of four (Ga,Mn)As layers with a thickness of 230 – 900 nm and the Mn content of 0.001%, 0.005%, 0.8% and 1.2%. The layers were grown by means of the LT-MBE method at a temperature of 230°C on semi-insulating (001)-oriented GaAs substrates. In addition, as a reference LT-GaAs layer, we investigated an 800-nm-thick undoped GaAs layer grown on GaAs by LT-MBE under the same conditions as the (Ga,Mn)As layers. Both the Mn composition and the layer thickness, which are listed in Table 1, were verified during growth by the reflection high-energy electron diffraction (RHEED) intensity oscillations.[24]

The layers were subjected to thorough investigations of their properties by means of a number of complementary characterization techniques. The structural properties of the layers were revealed from analysis of the results of high-resolution X-ray diffractometry (XRD) measurements performed at the temperature 27°C with the X-ray diffractometer equipped with a parabolic X-ray mirror and four-bounce Ge 220 monochromator at the incident beam and a three-bounce Ge analyzer at the diffracted beam. Misfit strain in the layers was investigated using reciprocal lattice mapping and rocking curve techniques for both the symmetric 004 and asymmetric 224 reflections of Cu $K\alpha_1$ radiation. Measurements of the sign of thermoelectric power were applied in order to determine the conductivity type in the epitaxial layers. Raman spectroscopy was employed to estimate the hole densities in the *p*-type (Ga,Mn)As layers. The micro-Raman measurements were performed using an inVia Reflex Raman microscope (Renishaw) at room temperature with the 532-nm semiconductor laser line as an excitation source. The magnetic properties of the (Ga,Mn)As layers were characterized with a superconducting quantum interference device (SQUID) magnetometer.

Modulation PR measurements were performed at room temperature using a semiconductor laser operating at the 532 nm wavelength and a nominal power of 20 mW, as a pump-beam source, and a 250 W halogen lamp coupled to a monochromator, as a probe-beam source. The PR signal was detected by a Si photodiode. The chopping frequency of the pump beam was 70 Hz and the nominal spot size of the pump and probe beams at the sample surface were 2 mm in diameter. Owing to the derivative nature of modulation PR spectroscopy it allows for accurate determination of the optical-transition energies even at room temperature.

## III. RESULTS OF THE LAYERS CHARACTERIZATION

High-resolution XRD measurements performed for all the layers revealed that both the LT-GaAs and (Ga,Mn)As epitaxial layers, grown on GaAs substrate under compressive misfit stress, were fully strained to the (001) GaAs substrate. The layers exhibited a high structural perfection, as proved by clear X-ray interference fringes revealed for the 004 Bragg reflections of all the layers, as shown in Fig. 1. The layer thicknesses calculated from the angular spacing of the fringes correspond very well to their thicknesses determined from the growth parameters. Diffraction peaks corresponding to the LT-GaAs and (Ga,Mn)As layers in Fig. 1 shift to smaller angles with respect to that of the GaAs substrate as a result of larger lattice parameters caused by incorporation of a large amount of excess arsenic in LT-GaAs and, additionally, Mn substitution atoms in (Ga,Mn)As. Angular positions of these diffraction peaks were used to calculate the perpendicular lattice parameters, $c$, the relaxed lattice parameters, $a_{rel}$, (assuming the (Ga,Mn)As elasticity constants to be the same as for GaAs), the lattice mismatch with respect to the GaAs substrate, and the vertical strain. The results, which confirm fairly well the Mn compositions of the (Ga,Mn)As layers determined from the RHEED measurements performed during the layers growth, are listed in Table 1. The lattice unit of the layers changes with increasing lattice mismatch from the zinc-blende cubic structure to the tetragonal structure with the perpendicular lattice parameter larger than the lateral one, equal to the GaAs substrate lattice parameter.

Measurements of the sign of thermoelectric power carried out for the investigated epitaxial layers have shown that both the LT-GaAs and the (Ga,Mn)As epitaxial layers with very low (0.001% and 0.005%) Mn contents were of $n$-type conductivity at room temperature. On the other hand, the (Ga,Mn)As layers with higher (0.8% and 1.2%) Mn contents displayed $p$-type conductivity.

The micro-Raman spectra for the LT-GaAs and (Ga,Mn)As layers are presented in Fig. 2. The spectra were recorded from the (001) surface in the backscattering configuration, where Raman scattering by longitudinal optical (LO) phonons via the deformation potential and Fröhlich electron-phonon interaction is allowed only. The signal of the symmetry-forbidden transverse optical (TO) phonons originates mainly from disorder scattering and is caused by a slight deviation from (001) surface orientation due to a residual strain in the epitaxial layers.[25] It is present in all the spectra shown in Fig. 2 and its magnitude increases with increasing Mn content and the mismatch strain in the layers, as shown in Table 1. The signal from LO phonons is clearly recognised for all the first-order micro-Raman spectra. Using two Lorentzian lines arising from the symmetry forbidden TO phonons and from the LO phonons we have performed a full line-shape analysis of the spectra measured for the LT-GaAs and the (Ga,Mn)As layers with very low (0.001% and 0.005%) Mn contents. However, such a decomposition of the spectra into two components was not sufficient for the results recorded for the (Ga,Mn)As layers with higher (0.8% and 1.2%) Mn contents, presented in Fig. 2.

In $p$-type GaAs with a high concentration of free holes the interaction between the hole plasmon and the LO phonon leads to the formation of Raman-active coupled plasmon–LO phonon (CPLP) mode.[26] Moreover, in heavily doped semiconductors with free carriers of very low mobility both the LO-phonon line and CPLP mode broaden and shift to the TO-phonon position with increasing carrier concentration.[26,27] Taking into account three contributions of Lorentzian shape, corresponding to the CPLP mode and the TO- and LO-phonon lines, we were able to fit the Raman spectra measured for the two (Ga,Mn)As layers with higher Mn contents of 0.8% and 1.2%, as shown in Fig. 2. Quantitative analysis of that type of Raman spectra can provide important information on the free-carrier density in heavily doped semiconductors. A reliable method of determining hole concentration in (Ga,Mn)As layers, based on analysis of their Raman spectra, without applying large magnetic fields, which is required in the Hall-effect measurements for ferromagnetic materials in order to extract the ordinary Hall effect from the dominating anomalous Hall effect, was proposed by Seong et al.[28] By examination of the relative Raman intensities of the LO-phonon and CPLP lines and their shift toward the TO-phonon position, and comparing them with the literature data,[27,28] we have estimated the hole concentrations of $0.5 \times 10^{19}$ cm$^{-3}$ and $8 \times 10^{19}$ cm$^{-3}$ in the (Ga,Mn)As layers with Mn contents of 0.8% and 1.2%, respectively.

SQUID magnetometry applied to all the investigated (Ga,Mn)As layers have revealed the ferromagnetic ordering for the layer with the highest, 1.2% Mn content only. The layer displayed an in-plane easy axis of magnetization, characteristic of compressively strained (Ga,Mn)As layers, and the Curie temperature of 60 K. The other layers, with lower Mn contents, displayed no measurable ferromagnetic ordering at temperatures down to 5 K. These results are in agreement with the assumed low Mn content and the estimated low concentration of free carriers in those layers.

## IV. PHOTOREFLECTANCE RESULTS AND DISCUSSION

The modulation PR spectra measured in the photon-energy range from 1.35 to 1.70 eV for both the LT-GaAs and (Ga,Mn)As epitaxial layers are shown in Fig. 3. The PR signal is associated with the electric field caused by the separation of photogenerated charge carriers. As a result of screening of this electric field by free carriers the intensity of measured PR signal strongly depends on free carrier concentration in the investigated layers. The spectra presented in Fig. 3 have been normalized to the same intensity. All the experimental spectra reveal a rich, modulated structure containing extrema around the band-gap-transition energy and the electric-field-induced Franz-Keldysh oscillations (FKOs) at energies above the fundamental absorption edge.[29] Moreover, the spectrum for the thinnest (Ga,Mn)As layer, of 1.2% Mn content and 230 nm thickness, exhibits an additional, below-band-gap feature, at the photon energy of about 1.42 eV. Despite the thicknesses of all the investigated epitaxial layers were larger than the penetration depth of 532-nm semiconductor laser pump beam in crystalline GaAs, estimated to about 150 nm, we have associated this feature with a contribution from the layer–substrate interface region.

The nature of such a feature, appearing on the low-energy side of the band-gap transition in PR spectra, was investigated in details by Sydor et al.[30] for differently doped GaAs epitaxial layers of various thicknesses grown by MBE on semi-insulating GaAs substrate. This feature, which increases with the free-carrier mobility and the doping of epitaxial layer, was observed for high-mobility GaAs layers of the thickness exceeding 1 μm, distorting their band-gap-transition PR signal. The feature was interpreted as a contribution to the PR signal from the layer–substrate interface region. According to this interpretation, it results from the transport of laser-beam-injected carriers to the GaAs/GaAs interface and formation of space-charge volume, which extends through the whole thickness of epitaxial layer. The modulation mechanism for this feature could come from the thermal excitation of impurities or traps at the interface and their momentary refilling by the laser-injected carriers.[30] In our previous investigations, with (Ga,Mn)As layers of higher (up to 6%) Mn content and thicknesses up to 300 nm, we observed the below-band-gap feature in the PR spectra of all the layers.[15]

The full-line-shape analysis of experimental PR spectra shown in Fig. 3, performed by using complex Airy functions and their derivatives,[31] allowed us to determine the energies of interband transitions and electro-optic energies for light and heavy holes in the investigated LT-GaAs and (Ga,Mn)As epitaxial layers. Because of the excitonic nature of the room-temperature PR signal, the measured energies for interband transitions in GaAs are smaller than the nominal band-gap energies by approximately the excitonic binding energy. The energies obtained from analysis of PR spectra are called critical-point energies $E_{CP}$. The critical-point energy corresponding to the fundamental band-gap transition at the Γ point of the Brillouin zone is denoted by $E_0$. The normalized change in reflection, $\Delta R/R$, measured in modulation PR spectroscopy can be defined as:[32]

$$\Delta R / R = \alpha \Delta \varepsilon_1 + \beta \Delta \varepsilon_2, \tag{1}$$

where $\alpha$ and $\beta$ are the Seraphin coefficients, and $\Delta\varepsilon_1$ and $\Delta\varepsilon_2$ are the pump-laser-induced changes in real and imaginary dielectric functions, respectively. In GaAs, near the $E_0$ critical point, the values of $\beta$ coefficient are very close to zero.[32] Consequently, the second term in Eq. (1) can be neglected in the fitting procedures and Eq. (1) takes the form: $\Delta R/R = \alpha\Delta\varepsilon_1$. The change in the real part of the dielectric function is expressed by:[31]

$$\Delta\varepsilon_1 = B\theta^{1/2} \operatorname{Im}\left[\frac{H(z)}{(\hbar\omega - i\gamma_0)^2}\right], \qquad (2)$$

where: $B$ is a constant related to the polarization and transition strength,

$$\hbar\theta = \left(\frac{e^2\hbar^2 F^2}{2\mu}\right)^{1/3} \qquad (3)$$

is the electro-optic energy (different for transitions from the light- and heavy-hole valence bands), where $e$ is the electron charge, $F$ is the electric field, $\mu$ is the interband reduced mass of the electron-hole pair for the corresponding transition, $\hbar\omega$ is the photon energy of the probe beam, $\gamma_0$ is a broadening parameter for the critical-point energy, and

$$H(z) = 2\pi\{\exp[(-\pi/e)i]A_i'(z)A_i'(w) + wA_i(z)A_i(w)\} - \left(\frac{-\eta + (\eta^2 + \gamma^2)^{1/2}}{2}\right)^{1/2} + i\left(\frac{\eta + (\eta^2 + \gamma^2)^{1/2}}{2}\right)^{1/2}$$

, where $\gamma = \gamma_0/\hbar\theta$ and $A_i$ and $A_i'$ are Airy functions and their derivatives, respectively. The arguments $z$ and $w$ of the Airy functions are given by the following formulas: $z = \eta + i\gamma = \frac{(E_{CP} - \hbar\omega)}{\hbar\theta} + i\frac{\gamma_0}{\hbar\theta}$ and $w = z\exp[(-2\pi/3)i]$.

In order to take into account both the light-hole (*lh*) and heavy-hole (*hh*) valence bands the experimental PR spectra were fitted using two corresponding Airy line-shape functions. In consequence, the relative changes in reflection coefficient were estimated using two following terms during the fitting procedure:

$$\Delta R/R = \alpha\Delta\varepsilon_1(lh) + \alpha\Delta\varepsilon_1(hh). \qquad (4)$$

In addition, for the thinnest (Ga,Mn)As layer (of 1.2% Mn content) an additional term in Eq. (4), described by the Aspnes third-derivative line-shape (TDLS) function,[33] was used in order to account for the below-band-gap feature in the PR spectrum related to a contribution from the layer–substrate interface region, as elaborated in our previous paper.[15] Calculations were done using the Levenberg-Marquardt algorithm with the Matlab 8.1 computer code. The functional form of the Seraphin coefficient $\alpha$ was applied. The fits are represented by solid lines in Fig. 3. The best fit parameters of $E_0$, $\hbar\theta_{lh}$ and $\hbar\theta_{hh}$ for all the layers studied are summarized Table 2. The $E_0$ value obtained for the LT-GaAs layer is in good agreement with earlier result of 1.42 eV obtained for LT-GaAs by Giordana et al. from their PR spectroscopy experiments.[34]

An alternative approach to analysis of the PR spectra containing FKOs consists in an examination of the energy positions of FKOs extrema in the PR spectra. This approach neglects the light-hole contribution to PR spectra and takes the asymptotic expressions of the Airy functions and their derivatives in Eq. (2).[35] As a result of this simplified analysis, which is, however, free from fitting parameters, the energy positions of FKOs extrema $E_m$ are linearly dependent on their effective index, defined as $F_m = [3\pi(m-1/2)4]^{2/3}$, according to the relation:[30]

$$E_m = E_G + \hbar\theta F_m, \tag{5}$$

where $m$ is the extremum number, the electro-optic energy $\hbar\theta$ is defined above by Eq. (3), and $E_G$ is the interband transition energy, closely related to the $E_0$ critical-point energy obtained from the full-line-shape analysis of PR spectra. The dependences of $E_m$ vs. $F_m$ for the investigated LT-GaAs and (Ga,Mn)As layers are plotted in Fig. 4. The values of $E_G$ and $\hbar\theta$, obtained from the intersection with ordinate and from the slope of this linear dependence, respectively, are listed in Table 2.

For all the layers the values of $E_G$ are smaller than those of $E_0$ but the sequence of their changes while increasing Mn content in the layers is very similar. On the other hand, the values of $\hbar\theta$, obtained from the analysis of FKOs periods, are slightly larger than those of $\hbar\theta_{hh}$, revealed from the full-line-shape analysis of the PR spectra, but the way they change with increasing Mn concentration is also the same. These similarities confirm correctness of the two approaches used in analysis of the measured PR spectra. Thorough inspection of the results presented in Table 2 shows that the $E_0$ critical-point energy, and also the $E_G$ energy, in (Ga,Mn)As layers decrease, with respect to those in the LT-GaAs layer, while increasing Mn content up to 0.005% and than increase with further increase of Mn content. For the (Ga,Mn)As layer with 1.2% Mn content both the $E_0$ and $E_G$ energies are larger than those in the LT-GaAs layer. In order to interpret these results we propose the band-structure evolution with increasing Mn concentration in (Ga,Mn)As layers with a low Mn content, schematically shown in Fig. 5.

In LT-GaAs the $As_{Ga}$ defects, owing to their high concentration, form $As_{Ga}$-related defect bands in the band gap responsible for hopping conduction[20] and the Fermi level pinning near the middle of the band gap.[20,36] This situation, resulting from partial ionization of the $As_{Ga}$-related defect band by residual carbon acceptors and/or native $V_{Ga}$ acceptors, is schematically shown in Fig. 5a. As a result of much larger dipole-transition strength for band-to-band optical transitions, than that for defect-to-band transitions, electronic transitions from the top of the valence band to the conduction band are predominant in the measured modulation PR spectra, as shown with an arrow in Fig. 5a.

In (Ga,Mn)As layers with very low Mn content and *n*-type conductivity the Fermi level position, determined by the compensation between the total density of incorporated acceptors, including the $Mn_{Ga}$ acceptors, and the $As_{Ga}$ donors, is still within the band gap. In order to account for the decrease in the interband transition energy with respect to that in LT-GaAs, revealed from our PR results, we assume a merging of the Mn-related impurity band with the GaAs valence band, as shown schematically in Fig. 5b, which refers to the (Ga,Mn)As layer with 0.005% Mn content. This assumption is in agreement with the results of recent four-wave mixing spectroscopy experiments by Yildirim et al.[37] who observed a strong increase in the optical response in the vicinity of the band gap, just for the (Ga,Mn)As layer with 0.005% Mn content, and attributed this observation to an increase in the density of states near the valence band edge caused by hybridization between the *d* levels of the $Mn_{Ga}$ and the *p* states of the host GaAs crystal. On the other hand, the maxima of their four-wave mixing spectra obtained for higher Mn contents, of 0.045% and 0.1%, shift to higher photon energies.[37]

Our PR results for the (Ga,Mn)As layers with higher, 0.8% and 1.2%, Mn content and *p*-type conductivity, showing an increase in the interband transition energy, are interpreted as a result of the Moss-Burstein shift of the absorption edge due to the Fermi level location within the valence band, determined by the free-hole concentration.[38] The band structure corresponding to the (Ga,Mn)As layer with 1.2% Mn content, displaying the Fermi level position below the top of the GaAs valence band, is schematically shown in Fig. 5c. These results are in agreement with the results of our earlier PR spectroscopy measurements for (Ga,Mn)As layers with Mn content of 1% and 2%.[15] On the other hand, in (Ga,Mn)As layers with much higher Mn content, of 4% and 6%, we revealed significant decrease in the band-gap-transition energy.[15] We interpreted those results in terms of a disordered valence band, formed as a result of merging the $Mn_{Ga}$-related impurity band with the host GaAs valence band, which extended within the band gap with increasing Mn content in (Ga,Mn)As layer.[15]

## V. CONCLUSIONS

We have studied the electronic- and band-structure properties of (Ga,Mn)As epitaxial layers with a low Mn content, in the range from 0 to 1.2% where the onset of ferromagnetic ordering occurs, by applying modulation photoreflectance spectroscopy and several complementary characterization techniques such as high-resolution X-ray diffractometry, thermoelectric power, Raman spectroscopy and SQUID magnetometery. PR spectroscopy results were elaborated by performing both the full-line-shape analysis of the PR spectra and the analysis of the periods of Franz-Keldysh oscillations, which concluded with similar findings on the evolution of the optical transition energies with increasing Mn content in the layers. Decrease in the band-gap-transition energy, with respect to that in the reference LT-GaAs layer, was revealed in very low-doped (Ga,Mn)As layer with Mn content of 0.001% - 0.005% and *n*-type conductivity. It is interpreted by assuming a merging of the Mn-related impurity band with the host GaAs valence band resulting in electronic transitions from the top of this disordered valence band to the conduction band. On the other hand, an increase in the band-gap-transition energy with increasing Mn content was observed in (Ga,Mn)As layers with higher Mn content of 0.8% and 1.2%, displaying *p*-type conductivity. It is interpreted as a result of the Moss-Burstein shift of the absorption edge due to the Fermi level position, determined by the free-hole concentration, within the valence band. The experimental results are consistent with the valence-band origin of mobile holes, which mediate ferromagnetic ordering in the (Ga,Mn)As diluted ferromagnetic semiconductor. The disordered character of the valence band may account for the observed very low mobility of holes in ferromagnetic (Ga,Mn)As layers.

## ACKNOWLEDGMENTS

O. Y. acknowledges financial support from the Foundation for Polish Science under Grant POMOST/2010-2/12 sponsored by the European Regional Development Fund, National Cohesion Strategy: Innovative Economy. This work was also supported by the Polish Ministry of Science and Higher Education under Grant No. N N202 129339. The MBE project at MAX-Lab is supported by the Swedish Research Council (VR).

Table 1. Layer thicknesses and their lattice parameters, $c$ and $a_{rel}$, lattice mismatch (defined as $(a_{rel} - a_{sub})/a_{sub}$, where $a_{sub} = 5.65349$ Å is the lattice constant of GaAs substrate), and vertical strain (defined as $(c - a_{rel})/a_{rel}$) for the investigated layers calculated from the results of high-resolution XRD measurements performed at 27°C.

| layer | thickness (nm) | $c$ (Å) (±0.00008) | $a_{rel}$ (Å) | lattice mismatch (×10$^4$) | vertical strain (×10$^4$) |
|---|---|---|---|---|---|
| LT-GaAs | 800 | 5.65527 | 5.65436 | 1.54 | 1.61 |
| 0.001% Mn | 900 | 5.65536 | 5.65440 | 1.61 | 1.70 |
| 0.005% Mn | 800 | 5.65546 | 5.65445 | 1.70 | 1.79 |
| 0.8% Mn | 300 | 5.65959 | 5.65646 | 5.25 | 5.53 |
| 1.2% Mn | 230 | 5.66509 | 5.65914 | 9.99 | 10.51 |

Table 2. The values of transition energies $E_0$ and electro-optic energies for light and heavy holes, $\hbar\theta_{lh}$ and $\hbar\theta_{hh}$, respectively, obtained from the full-line-shape analysis of the PR spectra shown in Fig. 3, and the values of energies $E_G$ and electro-optic energies $\hbar\theta$, obtained from the analysis of the FKOs period shown in Fig. 4.

| layer | $E_0$ (eV) | $\hbar\theta_{lh}$ (meV) | $\hbar\theta_{hh}$ (meV) | $E_G$ (eV) | $\hbar\theta$ (meV) |
|---|---|---|---|---|---|
| LT-GaAs | 1.423 | 25.91 | 22.01 | 1.415 | 23.64 |
| 0.001% Mn | 1.416 | 32.13 | 26.67 | 1.414 | 29.21 |
| 0.005% Mn | 1.407 | 52.60 | 42.32 | 1.395 | 44.35 |
| 0.8% Mn | 1.420 | 28.50 | 23.23 | 1.413 | 26.94 |
| 1.2% Mn | 1.429 | 50.12 | 38.30 | 1.417 | 42.69 |

**Figure captions**

Fig. 1. High-resolution X-ray diffraction results: $2\theta/\omega$ scans for (004) Bragg reflections for LT-GaAs and (Ga,Mn)As epitaxial layers grown on (001) semi-insulating GaAs substrate. The curves have been vertically offset for clarity. The narrow line corresponds to reflection from the GaAs substrate and the broader peaks at lower angles are reflections from the layers. With increasing Mn content the (Ga,Mn)As diffraction peaks shift to smaller angles with respect to that of the LT-GaAs reference layer.

Fig. 2. Raman spectra recorded at room temperature in backscattering configuration from the (001) surfaces of the LT-GaAs reference layer and four (Ga,Mn)As layers, where full symbols represent experimental data. The spectra have been normalized to the same intensity. The three lower spectra have been decomposed into two components of the TO- and LO-phonons, shown with solid lines. The two upper spectra have been decomposed into three components of the TO- and LO-phonons (solid lines) and the CPLP mode (cross-hatched area).

Fig. 3. Modulation photoreflectance spectra for the LT-GaAs reference layer and four (Ga,Mn)As layers with various Mn contents epitaxially grown on GaAs substrate (dots), where the FKOs extrema are marked with numbers. The spectra have been normalized to the same intensity and vertically offset for clarity. Solid lines represent fits to the experimental data by full-line-shape analysis of the spectra described in the text.

Fig. 4. Analysis of the period of Franz-Keldysh oscillations revealed in the PR spectra for the LT-GaAs reference layer and four (Ga,Mn)As layers with various Mn contents.

Fig. 5. Schematic energy band diagram for LT-GaAs (a) and its evolution for (Ga,Mn)As with increasing Mn content of 0.005% (b) and 1.2% (c). Splitting of the bands in the ferromagnetic state is omitted for simplicity. Arrows indicate electronic transitions from the valence band to the conduction band. $Mn_{Ga}$-related impurity band is assumed to be merged with the GaAs valence band in (b) and (c). In *p*-type (Ga,Mn)As the Fermi level position, determined by the free-hole concentration, lies within the valence band and the absorption edge shifts from the center of the Brillouin zone to the Fermi-wave vector, as shown in (c).

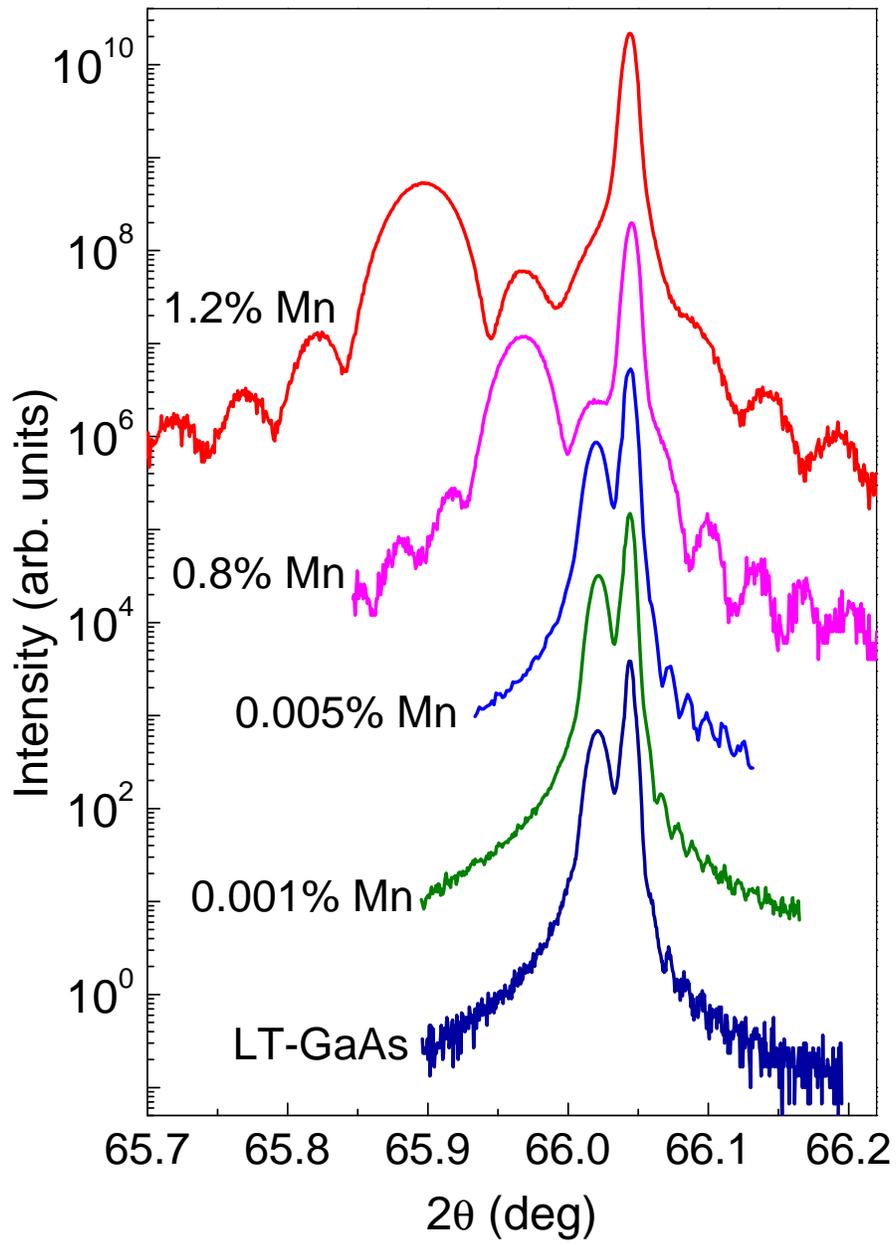

Fig. 1

Fig. 2

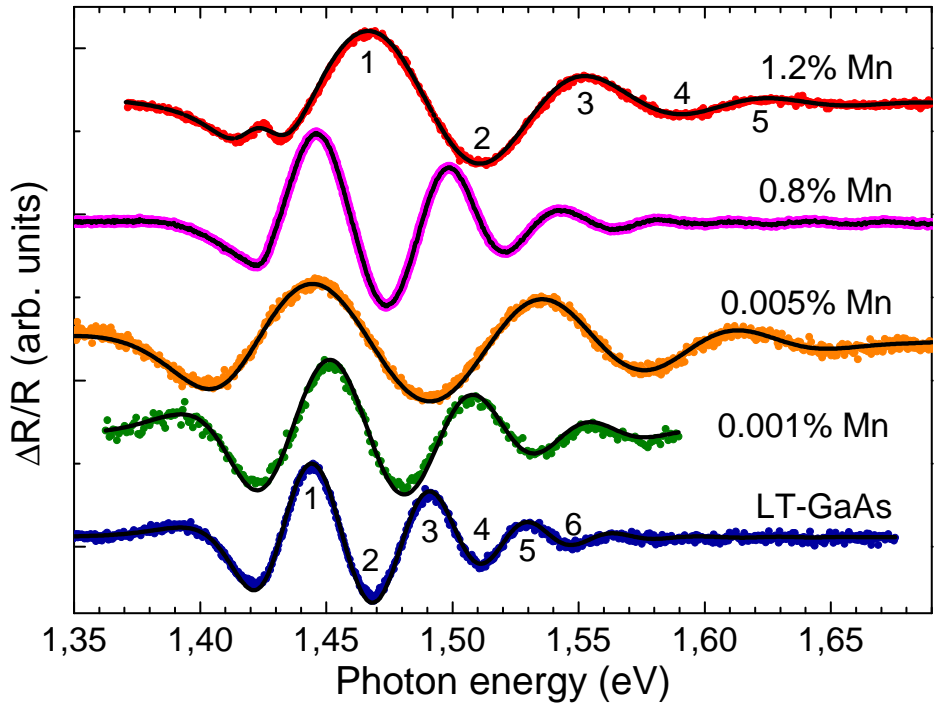

Fig. 3

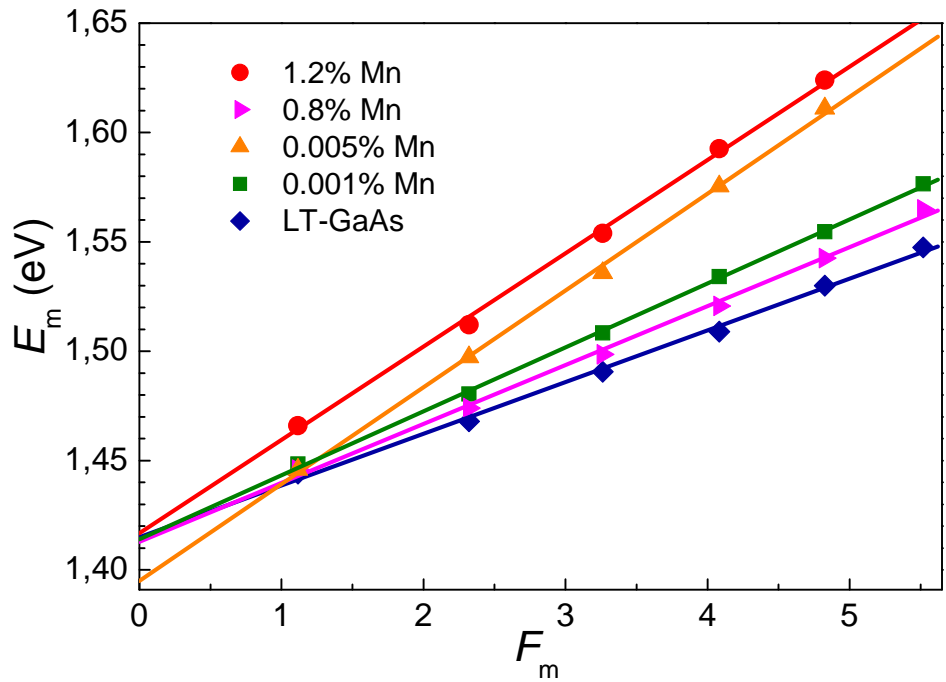

Fig. 4

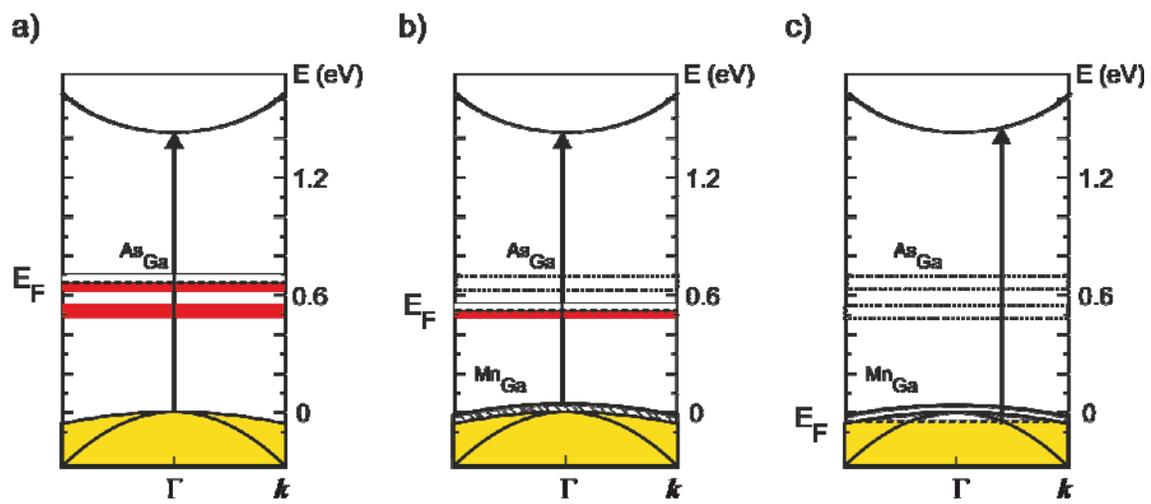

Fig. 5